\documentstyle[twocolumn,prb,aps,epsf]{revtex}

\begin{document}

{\bf Comment on ``Novel Convective Instabilities in a Magnetic
Fluid"}

\vskip 0.2 truecm

In a recent experiment [1], a vertical laser beam created a hot
spot in a thin horizontal ferrofluid layer. Since the Soret
coefficient for the fluid was positive and large, magnetic grains
(of which the ferrofluid was composed) effectively migrated
towards cold regions. Therefore, the concentration of the grains
inside the hot fluid volume appeared to be much lower than
outside, hence the hot spot became transparent. Being initially
round, the area of transparency transformed under the action of a
strong enough vertical magnetic field into $n$-lobe polygons. Luo,
Du, and Huang interpreted the phenomenon as ``a new convective
instability... The triangular shape originates from 6 convective
rolls with their axes parallel to the optic axis. Accordingly, an
8-roll state gives rise to a tetragon and 10-roll state a
pentagon". The rolls, however, exist only in imagination of the
authors of [1]. If the rolls had existed in reality, they would
have been directly visible in the same experiment. Fortunately,
there is no necessity in such an odd explanation based on {\em
unobservable} convective rolls. There is a much more plausible
interpretation of observations from [1].

The hot fluid volume may be considered as a nonmagnetic inset --
"bubble" -- surrounded by the magnetic fluid. The border of the
bubble has to be sufficiently distinct since the temperature
steeply decreases with the distance from the laser beam axis [1].
Thus one can introduce an effective radius $R$ of the hot droplet
and its surface tension $\sigma$ . The latter has a pure magnetic
origin: the volume fraction of magnetic grains $\phi$ was $6\%$ at
$r>R$, and it was negligible at $r<R$. Hence, the surface tension
on the border $r=R$ can be estimated by the expression $\sigma\sim
m^2/{\bar D}^5$, where $m$ stands for the magnetic moment of a
single particle and $\bar D$ is the mean distance between the
particles outside the droplet. Substituting $m=(\pi {\bar
d}^3/6)M_s$ and ${\bar D}={\bar d}\phi^{-1/3}$ we arrive at
$\sigma\simeq(\pi M_s/6)^2{\bar d}\phi^{5/3}$, where ${\bar
d}=9\,{\rm nm}$ is the mean particle diameter and $M_s=480\,{\rm
G}$ is the bulk magnetization of dispersed magnetite. For
afore-cited parameter values from [1], we obtain $\sigma\simeq
5.2\times 10^{-4}\,{\rm dyne/cm}$ what is {\em five orders} less
than the surface tension of water.
\begin{figure}
\begin{center}
\epsfxsize=7.0cm
\epsfbox{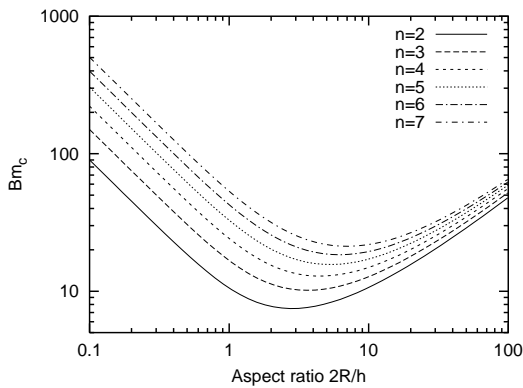}
\end{center}
\vspace*{-0.2cm} 
\caption{Critical magnetic Bond number as a function of droplet aspect ratio
for some modes of shape instability.}
\label{f1}
\end{figure}
Let us consider the inverse situation. As it is well-known, a
ferrofluid droplet placed in a horizontal Hele-Shaw cell
experiences under the vertical field $H$ the shape instability
[2,3]: the circle transforms into triangle, tetragon or other
$n$-lobe polygons in order to reduce its magnetic energy. The
competition between capillary and magnetic forces is characterized
by the magnetic Bond number $Bm=4M^2R/\sigma$ where $M=\chi H$ is
the fluid magnetization and $\chi$ the initial magnetic
susceptibility which obeys (approximately) the Curie law. The
critical value $Bm_c=(4\chi^2R/\sigma)H_c^2$ for the shape
instability with respect to a $n$-lobe mode depends on $n$ and on
the ratio $p=2R/h$ of the droplet diameter to the cell thickness
[4,5]. The dependence $Bm_c(p)$ for $n=2,\,3\,...,\,7$ calculated
by Drikis [6] is shown in Fig. 1.

Return now to the hot droplet. Luo, Du, and Huang reported that
``the temperature difference is about $15\,{\rm K}$ between the
beam axis and the beam edge. However, the temperature continues to
decrease outside the beam and levels off at {\em several beam
widths}$\,;$ the corresponding temperature difference is about
$40\,{\rm K}\,$''. So, as the radius of laser beam was
$\rho=6.7\,\mu{\rm m}$, an estimate $R\simeq 3\rho=20\,\mu{\rm m}$
looks pretty reasonable. The fluid used in [1] had the value
$4\pi\chi=0.94$ at room temperature. Thus we find $\chi=0.07$ at
$T=315\,{\rm K}$: by our estimate, it is the temperature at the
edge of the droplet, $r=R$. For the cell thickness $h=100\,\mu{\rm
m}$ (i.e., $p=0.4$) and $n=3$ (triangle) Fig. 1 predicts
$Bm_c=38$. Substituting above parameters into
$$
H_c=(2\chi)^{-1}\sqrt{(\sigma/R)Bm_c}\,,\eqno(1)
$$
we obtain $H_c=22.4\,{\rm Oe}$ that agrees well with experimental
value $H_c^{{\rm exp}}=22\,{\rm Oe}$. In experiment [1], the
critical field decreased when the sample thickness was lowered.
Indeed, for $h=50\,\mu{\rm m}$ and the same $R$ we have $p=0.8$
and $Bm_c=20$ which corresponds to the value $H_c=15.7\,{\rm Oe}$.
The latter is sufficiently close to $H_c^{{\rm exp}}=13\,{\rm
Oe}$. Thus, we have demonstrated that quite well established facts
of the physics of magnetic fluids are sufficient to explain the
experimental results [1] in a simple and reasonable way.

I am grateful to I. Drikis for providing Fig. 1 from [6].

\vskip0.3 truecm

Mark I. Shliomis

Department of Mechanical Engineering

Ben-Gurion University of the Negev

P.O.B. 653, Beer-Sheva, Israel

\vskip0.3 truecm




\vskip-0.2 truecm
\begin{references}
\bibitem{1} W. Luo, T. Du, and J. Huang, Phys. Rev. Lett. {\bf 82}, 4134 (1999).
\bibitem{2} E. Blums, A. Cebers, and M. Maiorov, {\em Magnetic Fluids} (W. de
Gruyter, Berlin, New York, 1997).
\bibitem{3} A.J. Dickstein, S. Erramilli, R.E. Goldstein, D.P. Jackson, and S.A.
Langer, Science {\bf 261}, 1012 (1993).
\bibitem{4} D.P. Jackson, R.E. Goldstein, and A. Cebers, Phys. Rev. E {\bf 50},
298 (1994).
\bibitem{5} J.A. Miranda and M. Widom, Phys. Rev. E {\bf 55},
3758 (1997).
\bibitem{6} I. Drikis, {\em Nonlinear dynamics of magnetic fluid free surface in
Hele-Shaw cell} (PhD Thesis, University Paris 7, 1999).
\end{references}
\end{document}